\documentclass[a4paper,english,12pt]{article}
\usepackage[T1]{fontenc}
\usepackage[latin1]{inputenc}
\usepackage{babel}
\usepackage{color}
\usepackage{a4wide}
\usepackage{hyperref}
\usepackage{graphicx}
\usepackage{amsmath}
\usepackage{amssymb}
\begin{document}

\title{\textbf{Minimal violation of flavour and custodial symmetries in a vectophobic Two-Higgs-Doublet-Model}}
\author{Elvira Cerver\'{o}\footnote{elvira.cervero@uclouvain.be}\hspace{0.3cm}and Jean-Marc G\'{e}rard\footnote{jean-marc.gerard@uclouvain.be}}
\date{}
\maketitle
\begin{center}
\textit{Centre for Cosmology, Particle Physics and Phenomenology (CP3),\\
Universit\'{e} catholique de Louvain,\\
Chemin du Cyclotron 2, B-1348, Louvain-la-Neuve, Belgium.
}
\end{center}
\begin{abstract}
Tree-level accidental symmetries are known to play a fundamental role in the phenomenology of the Standard Model (SM) for electroweak interactions. So far, no significant deviations from the theory have been observed in precision, flavour and collider physics. Consequently, these global symmetries are expected to remain quite efficient in any attempt beyond the SM. Yet, they do not forbid rather unorthodox phenomena within the reach of current LHC experiments. This is illustrated with a vectophobic Two-Higgs-Doublet-Model (2HDM) where effects of a light, flavour-violating and custodian (pseudo)scalar might be observed in the $B_s\to\mu^+\mu^-$ decay rate and in the diphoton invariant mass spectrum at around 125 GeV.
\end{abstract}

\section{Introduction}

Baryon number conservation, invoked \cite{Wigner} to explain the striking stability of the proton against $p\to e^+\gamma$ , played a crucial role in the building of the quark model and turned out to be a tree-level accidental symmetry of the SM. Indeed, its associated $U(1)_B$  group is only broken by very tiny quantum effects linked to the gauge coupling of $SU(2)_L$. Remarkably, once splitted into distinct sectors, the $SU(2)_L \times U(1)_Y$ gauge-invariant SM Lagrangian has progressively revealed other accidental global symmetries that are now quite useful for our understanding of electroweak processes among the three up and down quarks.

On the one hand, the so-called custodial symmetry is an accidental one arising from the Higgs potential of the SM. It has been identified \cite{Sikivie:1980hm} as the responsible for the amazing success of the tree-level mass relation
\begin{equation}
\rho\equiv\frac{M^2_W}{M_Z^2\cos^2{\theta_W}} = 1
\end{equation} with respect to the electroweak precision data. Indeed, its associated $SU(2)_{L+R}$ group is only explicitly broken by the up-down quark mass splittings, in the limit where the gauge coupling of $U(1)_Y$ can be neglected (or equivalently if $\theta_W\to0$).

On the other hand, the large flavour symmetry \cite{Chivukula:1987py} with unitary transformations acting respectively on the left-handed quark doublets, the right-handed charge $\frac{2}{3}$ quarks and the right-handed charge $-\frac{1}{3}$ quarks is an accidental symmetry in the Yang-Mills sector of the SM. It is used to classify all Flavour Changing Neutral Currents (FCNC) beyond tree-level in terms of the Cabibbo-Kobayashi-Maskawa (CKM) mixing matrix
\begin{equation}
V_{CKM}=U^u_LU^{d\dag}_L
\end{equation}
\noindent where the $U^{u,d}_L$ are misaligned unitary matrices relating the $u_L$ and $d_L$ weak eigenstates to their mass eigenstates. In other words, the CKM matrix is obviously invariant under the associated $U(3)_{Q_L} \times U(3)_{U_R} \times U(3)_{D_R}$ but these flavour groups are explicitly broken by the up-up (down-down) quark mass splittings.

In the SM, both the accidental (bosonic) custodial and (fermionic) flavour symmetries are violated by the Yukawa couplings of the single Higgs field to the quark ones, in a way consistent with all the available data. Naively, the safest way to go beyond the SM is to ensure a minimal violation of these global symmetries. Yet, this does not necessarily guarantee orthodoxy. Indeed, such an extension of the SM through the introduction of a second (vectophobic) Higgs doublet might already lead to non-standard $B_s\to\mu^+\mu^-$ decay rate and two-photon invariant mass spectrum at running LHC experiments. 

\section{Custodial symmetry}

If we impose a custodial symmetry on the 2HDM potential, the physical states can be naturally classified in triplet and singlet irreducible representations of the unbroken $SU(2)_{L+R}$, namely
\begin{equation}
\Phi_1 \ni\left\{\begin{array}{ccc}G^{+}\\G^0\\G^-\end{array}\right\}\oplus\{h^0+\frac{v}{\sqrt 2}\};\qquad v=(\sqrt{2}G_F)^{-\frac{1}{2}}\approx 246 \mbox{ GeV}
\label{phi1}
\end{equation}
and
\begin{equation}
\Phi_2 \ni\left\{\begin{array}{ccc}H^{+}\\A^0\\H^-\end{array}\right\}\oplus \{H^0\}\mbox{  or  }\left\{\begin{array}{ccc}H^{+}\\H^0\\H^-\end{array}\right\}\oplus \{A^0\}.\label{phi2}
\end{equation}
Retaining CP violation as part of flavour violation, we assume the spin-0 sector to be CP invariant with $H^0$ and $A^0$ the new scalar and pseudoscalar, respectively. Based on a custodial symmetry, these assignments with $h^0$ behaving as the SM scalar and all the new physical states beyond the SM being in $\Phi_2$ would correspond to a particular case of the so-called Higgs basis  \cite{Branco:2011iw} (defined by the angle $\beta$ for the $G^{\pm}$-$H^{\pm}$ and $G^0$-$A^0$ mixings) with a further assumption on the $H^0$-$h^0$ mixing angle, namely $\alpha=\beta-\frac{\pi}{2}$. The triplet in eq. \eqref{phi1} corresponds to the massless Nambu-Goldstone bosons. In the limit where the scalar triplet in eq. \eqref{phi2} is also degenerate in mass, the custodial $SU(2)_{L+R}$ symmetry is minimally broken like in the SM (i.e., by $m_b\ll m_t$ and $\theta_W\neq 0$) since $\Phi_2$ is vectophobic and its quantum corrections to the $\rho$ parameter cancel. In the following, we will assume that the singlet component of $\Phi_2$ is light compared to its triplet partners, i.e.
\begin{equation}
m_{H^0} < m_{A^0} \approx m_{H^{\pm}}\hspace{0.3cm}\mbox{or}\hspace{0.3cm} m_{A^0} < m_{H^0} \approx m_{H^{\pm}},   \label{hierarchies}                     
\end{equation}                
\noindent to seek out new physics beyond the SM in current experiments. As for the SM-like $h^0$ mass, it remains temporarily a free parameter depending on the triplet mass splitting  through the $\rho$ parameter. Note that the second, CP-twisted, case with a light pseudoscalar $A^0$ may also naturally arise either from a spontaneous symmetry breaking \cite{Gerard:2007kn,deVisscher:2009zb} or from a dynamical \cite{Burdman:2011fw} one. Yet, the option of a fundamental or effective Higgs potential is left open hereafter.

\section{Flavour symmetries}

In the custodial 2HDM characterized by eqs. \eqref{phi1} and \eqref{phi2}, the Yukawa couplings are given by
\begin{equation}
\mathcal{L}_Y = - \bar Q'_L\left(Y'_d\Phi_1+Z'_d\Phi_2\right)d'_R-\bar Q'_L\left(Y'_u\tilde{\Phi}_1+Z'_u\tilde{\Phi}_2\right)u'_R + h.c.\label{2hdmlagr}
\end{equation}

\noindent Consequently, all the fermions acquire a mass through their coupling $Y'$ to $\Phi_1$ while tree-level FCNC are induced by their coupling $Z'$ to the new spin-0 fields in $\Phi_2$. As these FCNC are very much constrained by experimental data, some mechanism must be found in order to suppress them. A popular way to forbid any FCNC at tree-level is the so-called Natural Flavour Conservation (NFC) hypothesis \cite{Glashow:1976nt} based on a flavour blind symmetry. However, here, if the Higgs doublets have a different parity under the discrete group $\mathbb{Z}_2$ ($\Phi_1\to\Phi_1$ and $\Phi_2\to-\Phi_2$) the $Z'$ couplings are not allowed, the vectophobic $\Phi_2$ becomes also fermiophobic and we simply recover the flavour physics of the SM. 

So, let us consider another way to tame but not eliminate FCNC at tree-level beyond the SM, namely the Minimal Flavour Violation (MFV) hypothesis \cite{Chivukula:1987py,D'Ambrosio:2002ex,Isidori:2012ts,Buras:2010mh,Botella:2009pq}. To formulate the MFV hypothesis, one first considers the full flavour symmetry of the gauge sector. Although the Yukawa couplings explicitly break $G_f=SU(3)_{Q_L}\times SU(3)_{U_R}\times SU(3)_{D_R}$, this symmetry can be restored by imposing suitable transformation laws under $G_f$ to them,
\begin{eqnarray}
			Y'_u&\sim&(3,\bar{3},1)\label{Yu}\\
			Y'_d&\sim&(3,1,\bar{3}).\label{Yd}
\end{eqnarray}
By doing so, the Yukawa couplings are promoted into auxiliary fields or spurions. The MFV hypothesis as formulated in \cite{Smith:2009hj} and implemented here is based on two conditions:
\begin{itemize}
\item The new flavour structures beyond the SM must be invariant under the $G_f$ group. To implement this first condition, the new flavour structures are written as a series in terms of the spurions. The minimality of the hypothesis is guaranteed by imposing that only the spurions needed to account for the fermion masses and mixings are allowed.
\item The coefficients of the MFV expansion in terms of the spurions must be natural, i.e. $\mathcal{O}(1)$. This second condition is imposed to let the spurions be the only structures responsible for the masses and mixing hierarchies and to avoid any further fine-tuning.
\end{itemize}

The three $U(1)$ symmetries forsaken in eqs. \eqref{Yu} and \eqref{Yd} can be rearranged to correspond respectively to the vectorial baryon number, the chiral hypercharge and the axial Peccei-Quinn charge,
\begin{equation}
U(1)^3= U(1)_B\times U(1)_Y\times U(1)_{PQ}.
\end{equation}
In the SM, the $U(1)_B$ remains an accidental symmetry while $U(1)_Y$ is spontaneously broken since the Higgs doublet carries no baryon number but a non-zero hypercharge. In a general 2HDM, it is also possible to decouple the breaking of the $U(1)_{PQ}$ by shifting the PQ charge of the spurions to the Higgs doublets. However, in eq. \eqref{2hdmlagr} only $\Phi_1$  generates the quark masses and mixings such that the minimality requirement would then imply massless up or down quarks.

In the past, spurions were introduced for a straightforward isospin decomposition of the weak $K\to\pi\pi$ decay amplitudes or to provide Goldstone bosons with a small mass in a chiral invariant effective theory for strong interactions. In the first case, these spurions are just auxiliary fields with no physical meaning while in the second one, the light quark mass matrix promoted to a field is eventually related to the Higgs one. Here, MFV gives rise to an effective low-energy theory which does not make any assumption about the possible underlying high-energy dynamics of the spurions. They could well be the background values of new heavy scalar fields called flavons \cite{Alonso:2011yg}.

To apply the specific formulation of MFV given above to eq. \eqref{2hdmlagr}, we simply have to express the new flavour structures $Z'_i$ as series of the $Y'_i$ couplings in a $G_f$ invariant way. If we neglect down quark masses with respect to the top one, the $Y'_d$ coupling can be set to zero inside the series. Using the Cayley-Hamilton relation for a $3\times3$ Hermitian matrix, we then obtain the following Yukawa couplings to $\Phi_2$
	 \begin{eqnarray}
	 Z'_d &=& \{\delta_0  + \delta_1Y'_uY'^{\dag}_u + \delta_2(Y'_uY'^{\dag}_u)^2\}Y'_d,\label{ZdMFV}\\
	 Z'_u &=& \{\upsilon_0 + \upsilon_1Y'_uY'^{\dag}_u + \upsilon_2(Y'_uY'^{\dag}_u)^2\}Y'_u.\label{ZuMFV}
	 \end{eqnarray}

In the aligned 2HDM \cite{Pich:2009sp, Jung:2010ik}, the relations among the Yukawa couplings are equivalent to eqs. \eqref{ZdMFV} and \eqref{ZuMFV} when considering only $\delta_0$ and $\upsilon_0$ and allowing them to be complex. Yet, the other terms are all assumed to be induced through quantum effects such that $\delta_{1,2}$ and $\upsilon_{1,2}$ are loop factors much smaller than one. In this limit, NFC is recovered and relations to the $\tan\beta$ ($\cot\beta$) coefficients for the various $\mathbb{Z}_2$ invariant Type-I and Type-II models can easily be established. The contribution to the $\bar B_s\to X_s\gamma$ of the 2HDM analysed in this work does not differ much from the one in \cite{Degrassi:2010ne}. So, the resulting limit imposed on the charged Higgs mass is
\begin{equation}
m_{H^{\pm}}\gtrsim400\mbox{ GeV}
\end{equation}

\noindent for natural values of the MFV coefficients, namely $\delta_i$ and $\upsilon_i$ close to one. Hereafter, we will saturate this bound by taking the mass of $H^{\pm}$ and its custodial neutral partner (see eq. \eqref{hierarchies}) at 400 GeV. Note that the new LHC bounds do not apply to the heavy $H^0(A^0)$ since it may decay in a non-standard way via $H^0(A^0)\to A^0(H^0)Z^0$.

The MFV hypothesis has already been implemented in various 2HDM. In \cite{D'Ambrosio:2002ex,Isidori:2012ts}, the $SU(2)_L\times U(1)_Y$ gauge-invariant operators were classified. These dimension-six operators are $\Lambda^{-2}$ suppressed only if the scale hierarchy $M_W\ll m_{H^0,A^0}\ll\Lambda$ is assumed. Such an effective approach differs from ours since we consider MFV with $H^0$ \underline{or} $A^0$ lighter than the top quark and rather close to the W-mass scale. In \cite{Buras:2010mh}, the MFV hypothesis was formulated in the generic basis with CP violation in a Type-II model for large $\tan\beta$. Here, by formulating MFV directly in a CP-invariant and vectophobic basis, we have simply rotated away any $\tan\beta$ dependence.

In our vectophobic 2HDM, the down tree-level FCNC are induced by the following Yukawa interactions expressed in terms of the quark mass eigenstates
\begin{equation}
\mathcal{L}^{FCNC}_Y=-\bar{d}^i_L(Z_d)_{ij}d^j_R\left(\frac{H^0 +iA^0}{\sqrt{2}}\right)+h.c.\label{dFCNC}
\end{equation}

\noindent with $i\neq j$,

\begin{equation}
Z_d=4G_F\delta_1V^{\dag}_{CKM}M^2_uV_{CKM}\frac{M_d}{v},\label{ZdCKM}
\end{equation}

\noindent and $M_{u(d)}$, the diagonal up (down) quark mass matrix. Note that we have only taken into account the first non diagonal term in eq. \eqref{ZdMFV}. Indeed, the huge mass hierarchy in the up sector implies that $(Y^{\dag}_uY_u)^2$ is almost aligned to $Y^{\dag}_uY_u$ in the three dimensional flavour space and the naturalness principle imposes that the  $\delta_i$ and the $\upsilon_i$ are $\mathcal{O}(1)$. If we only consider the leading contribution from the top quark mass, the $Z_d$ coupling can be expressed like
\begin{equation}
(Z_d)_{ij}=4G_F\delta_1(V^*_{ti}V_{tj})m^2_t\frac{m_{d_j}}{v}.\label{Zd}
\end{equation}
As far as the up tree-level FCNC are concerned, they are simply absent from eq. \eqref{ZuMFV}. This implies that we will not consider the $D^0$ meson mixing and decays which are anyway polluted by sizeable long-distance effects.

\subsection{$\Delta F = 2$ mixings}

In the following, we analyse the implications of our custodial 2HDM with MFV in a few $\Delta F = 2$ quantities. In particular, new contributions to the $B_s$ meson mass difference as well as to the $|\epsilon_K|$ parameter that estimates the amount of CP violation in the neutral Kaon system will be studied. In the SM, these quantities are dominated by short distance (SD) transitions and mainly induced through virtual top quark box diagrams.

The $\Delta F =2$ effective Hamiltonian associated to the Yukawa interactions given in eq. \eqref{dFCNC} reads
\begin{equation}
\mathcal{H}^{\Delta F=2}_{2HDM}=\left(-\frac{1}{m^2_{H^0}}\frac{[(Z_d)_{ij}-(Z^{\dag}_d)_{ij}]^2}{2}+\frac{1}{m^2_{A^0}}\frac{[(Z_d)_{ij}+(Z^{\dag}_d)_{ij}]^2}{2}\right)(\bar d^i_{R}d^j_{L})(\bar d^i_{R}d^j_{L}).\label{hamiltonianMFV}
\end{equation}

\noindent Within the SM, the SD $\Delta F=2$ transitions take place through the one-loop box diagrams and the corresponding effective Hamiltonian is proportional to the operator $\mathcal{O}_{SM}=(\bar d_L^i\gamma^{\mu}d_L^j)(\bar d_L^i\gamma_{\mu}d_L^j)$. At the hadronic level, using eq. \eqref{Zd} and with the help of the Dirac equation, one can thus express the matrix element for eq. \eqref{hamiltonianMFV} as follows
\begin{equation}
\langle \bar M^0|\mathcal{H}^{\Delta F=2}_{2HDM}|M^0\rangle\simeq\frac{8 G^2_F\delta_1^2(V^*_{ti}V_{tj})^2m^4_tm^2_M}{v^2}\left[\frac{(m_{d_i}-m_{d_j})^2}{(m_{d_i}+m_{d_j})^2}\frac{1}{m^2_{H^0}}-\frac{1}{m^2_{A^0}}\right]\langle \bar M^0|\mathcal{O}_{SM}|M^0\rangle.\label{hamiltonianMFV2}
\end{equation}

\noindent For $K^0$ $(\bar s d)$ and  $B^0_q$ $(\bar b q)$ mesons, the down quark mass hierarchy ($m_d\ll m_s\ll m_b$) allows us to consider the limit $m_{d_j}\ll m_{d_i}$ and we get
\begin{equation}
\langle \bar M^0|\mathcal H^{\Delta F =2}_{eff}|M^0\rangle\simeq\langle \bar M^0|\mathcal H^{\Delta F =2}_{eff}|M^0\rangle^{SM}\left[1+16\pi^2 x \delta^2_1 m^2_M\left(\frac{1}{m^2_{H^0}}-\frac{1}{m^2_{A^0}}\right)\right].
\label{matrixelement}\end{equation}

\noindent In eq. \eqref{matrixelement}, the factor $16\pi^2$ stems from the SM one-loop contribution while $x$ encodes the full dependence on the top quark mass ($m_t(m_t)=163.4$ GeV)
\begin{equation}
x=\frac{2m_t^4}{M^2_Wv^2S_0(x_t)}\approx 1.61.
\end{equation}

The sign of the CP-odd ($-$) $A^0$ and CP-even (+) $H^0$ exchange contributions in eq. \eqref{matrixelement} can easily be understood by comparing with the general expression for any long-distance (LD) contributions to a $\Delta F = 2$ transition \cite{Marshak}
\begin{equation} 
\langle\bar{M^0}|\mathcal H^{\Delta F =2}_{eff}|M^0\rangle_{LD} = \sum_{I} \left(\frac{|\langle M(-)|\mathcal H^{\Delta F =1}|I(-)\rangle|^2}{m_M-E_{I(-)} + i\epsilon}-\frac{|\langle M(+)|\mathcal H^{\Delta F =1}|I(+)\rangle|^2}{m_M-E_{I(+)} + i\epsilon}\right),
\end{equation}

\noindent in the limit of CP-invariance. For illustration, in the case of the $K_L$-$K_S$ mass difference, the single pseudoscalar exchange LD contribution is positive for the pion (lighter than the kaon) but negative for the $\eta^{(')}$ (heavier than the kaon) \cite{Donoghue:1983hi}, leading to a rather strong cancellation in the chiral perturbation theory \cite{Gerard:2005yk}.

\begin{figure}[htbp]
\begin{tabular}{cc}
\includegraphics[scale=0.26]{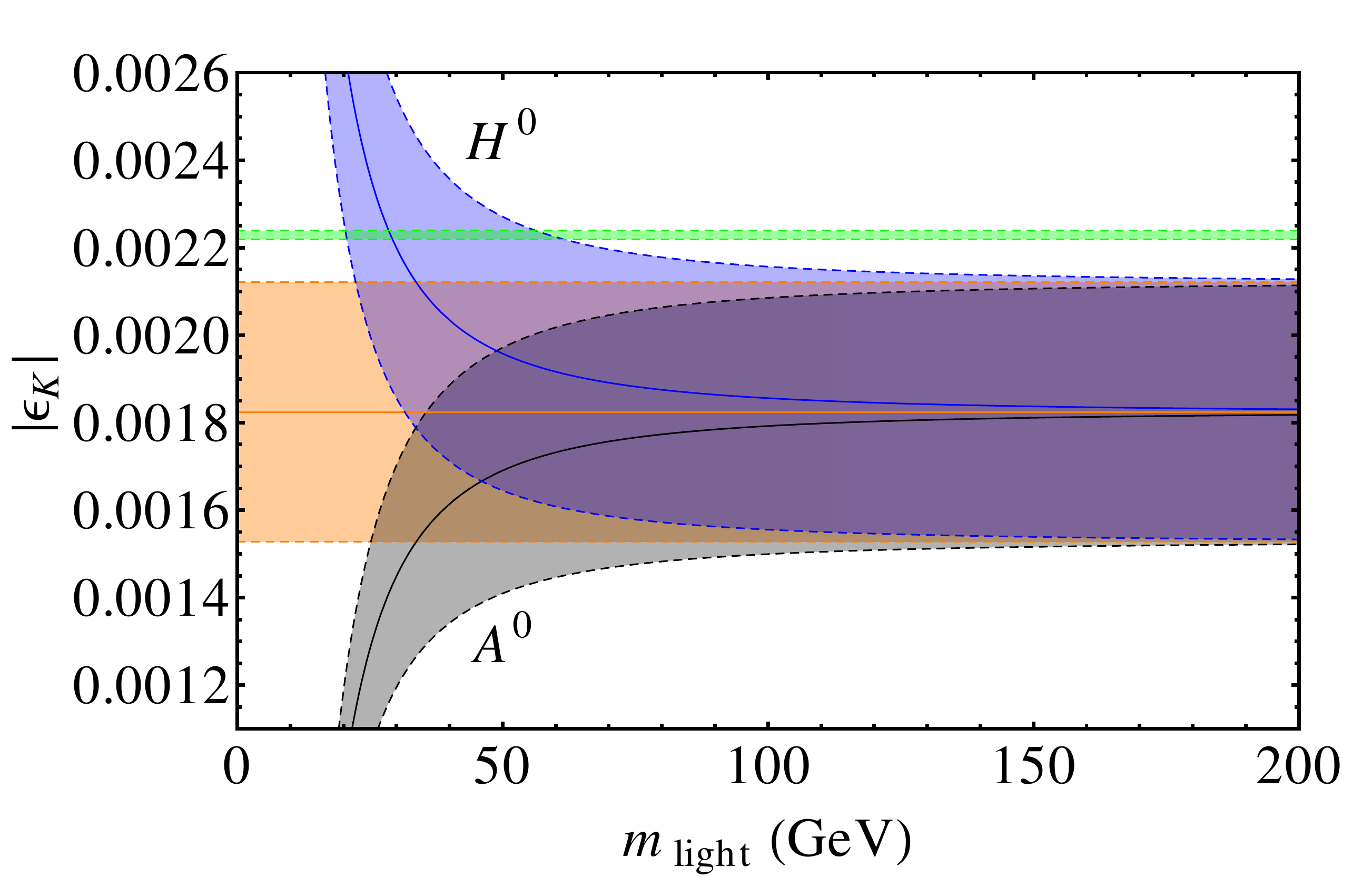}		&\includegraphics[scale=0.26]{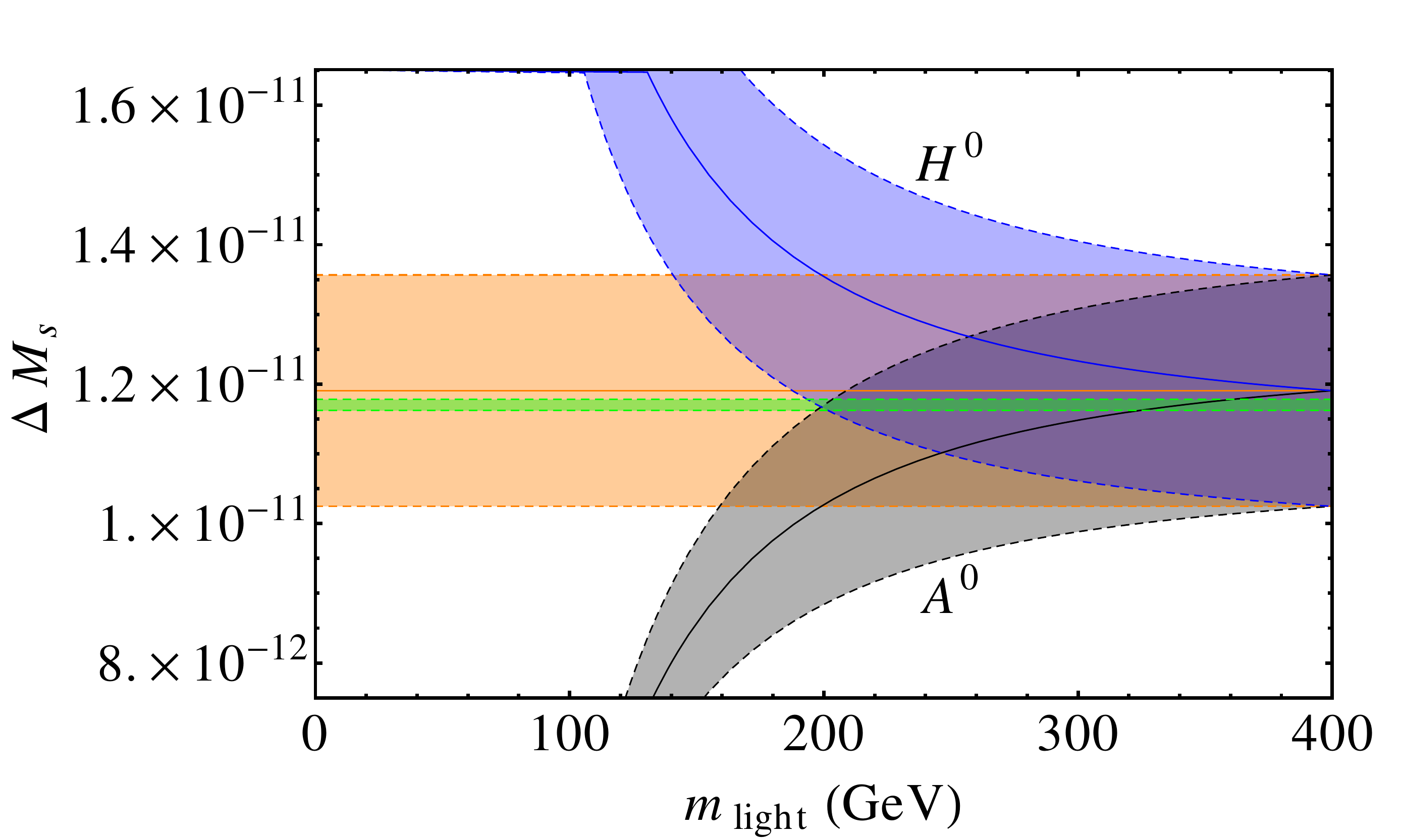}
\end{tabular}
 \caption{\small{\textit{$|\epsilon_K|$ and $\Delta M_{B_s}$ as a function  of the $H^0$ and $A^0$ masses for a MFV coefficient $\delta_1=1$. The thin horizontal (green) bands indicate the experimental values, the broad horizontal (orange) bands indicate the 1$\sigma$ SM prediction, the upper (blue) curved bands show the 1$\sigma$ prediction for $m_{H^0}\ll m_{A^0}=400$ GeV, while the lower (grey) curved bands correspond to the analogue prediction for $m_{A^0}\ll m_{H^0}=400$ GeV. The theoretical values of the SM and the experimental ones are given in Tab.\ref{numerics2}.}}}\label{neutralmesons}\end{figure}

The measured $|\epsilon_K|$ parameter in the $K^0$-$\bar K^0$ system would alone clearly welcome some enhancement, i.e. a rather light $H^0$, to relax a potential tension within the SM \cite{Lunghi:2008aa,Buras:2008nn}. However, as displayed in figure \ref{neutralmesons}, the $\bar B_s$-$B_s$ system already excludes such a scenario. In fact, eq. \eqref{matrixelement} directly tells us what will be the effect of a light $H^0$ or $A^0$ on the various $\Delta F=2$ systems. Indeed, the new term  with respect to the SM contribution is proportional to the square of the meson mass $m_M$. Such a feature implies a bigger effect in the B-meson system case than in the Kaon one and almost no difference between the $B_d$ and $B_s$ systems. Contrariwise, in the decoupling limit with $m_{A^0}=m_{H^0}\approx\Lambda$, eq. \eqref{hamiltonianMFV2} tells us that any correction with respect to the SM should scale as $\left(\frac{m_{d_j}}{m_{d_i}}\right)\left(\frac{m^2_M}{\Lambda^2}\right)\ll1$.\\

\subsection{The $B_s\to\mu^+\mu^-$ decay}

In the SM, the rare $B_s\to\mu^+\mu^-$ decay takes place through box and penguin diagrams, leading to an effective Hamiltonian proportional to the single axial operator
\begin{equation}
Q_A=(\bar b_L\gamma^{\mu}s_L)(\bar{\mu}\gamma_{\mu}\gamma_5\mu).\label{VA}
\end{equation}

\noindent  However, when introducing new physics beyond the SM, contributions from other operators can be sizeable. In the specific 2HDM under scrutiny, tree-level flavour changing neutral Higgs exchanges induce the following new scalar and pseudoscalar operators \cite{Logan:2000iv}
\begin{eqnarray}
Q_S&=&m_b(\bar b_Rs_L)(\bar{\mu}\mu)\label{scalar}\\
Q_P&=&m_b(\bar b_Rs_L)(\bar{\mu}\gamma_5\mu),\label{pseudoscalar}
\end{eqnarray}

\noindent if the limit $m_s\ll m_b$ is taken again. In order to compute the branching ratio associated to this decay, MFV needs to be introduced in the lepton sector as well. This can be done in analogy with the quark sector. Yet, the specific lepton mass spectrum allows us to truncate the MFV series for $Z_{\ell}$ at first order
\begin{equation}
Z_{\ell}=\lambda_0Y_{\ell}.
\end{equation}

In any model where the operators in eqs. \eqref{scalar} and \eqref{pseudoscalar} give non-negligible contributions, the branching ratio can be expressed as follows
\begin{equation}
\mathcal{B}(B_s\to\mu^+\mu^-)= \mathcal{B}(B_s\to\mu^+\mu^-)^{SM}\left[\left(1+m^2_{B_s}\frac{C_P}{C_A}\right)^2+\left(1-\frac{4m^2_{\mu}}{m^2_{B_s}}\right)m^4_{B_s}\frac{C_S^2}{C^2_A}\right]\label{branching}
\end{equation}

\noindent where $C_A=2Y(x_t)\approx 2.0$ is the Wilson coefficient associated to the SM axial operator. In our 2HDM, the coefficients $C_S$ and $C_P$ are defined by
\begin{equation}
C_{S(P)}=\frac{\Delta}{m^ 2_{H^0(A^0)}};\qquad \Delta = \frac{4\pi^2 \delta_1\lambda_0m^2_t}{M^2_W}.
\end{equation}

\begin{figure}[htbp]
\centering\includegraphics[scale=0.31]{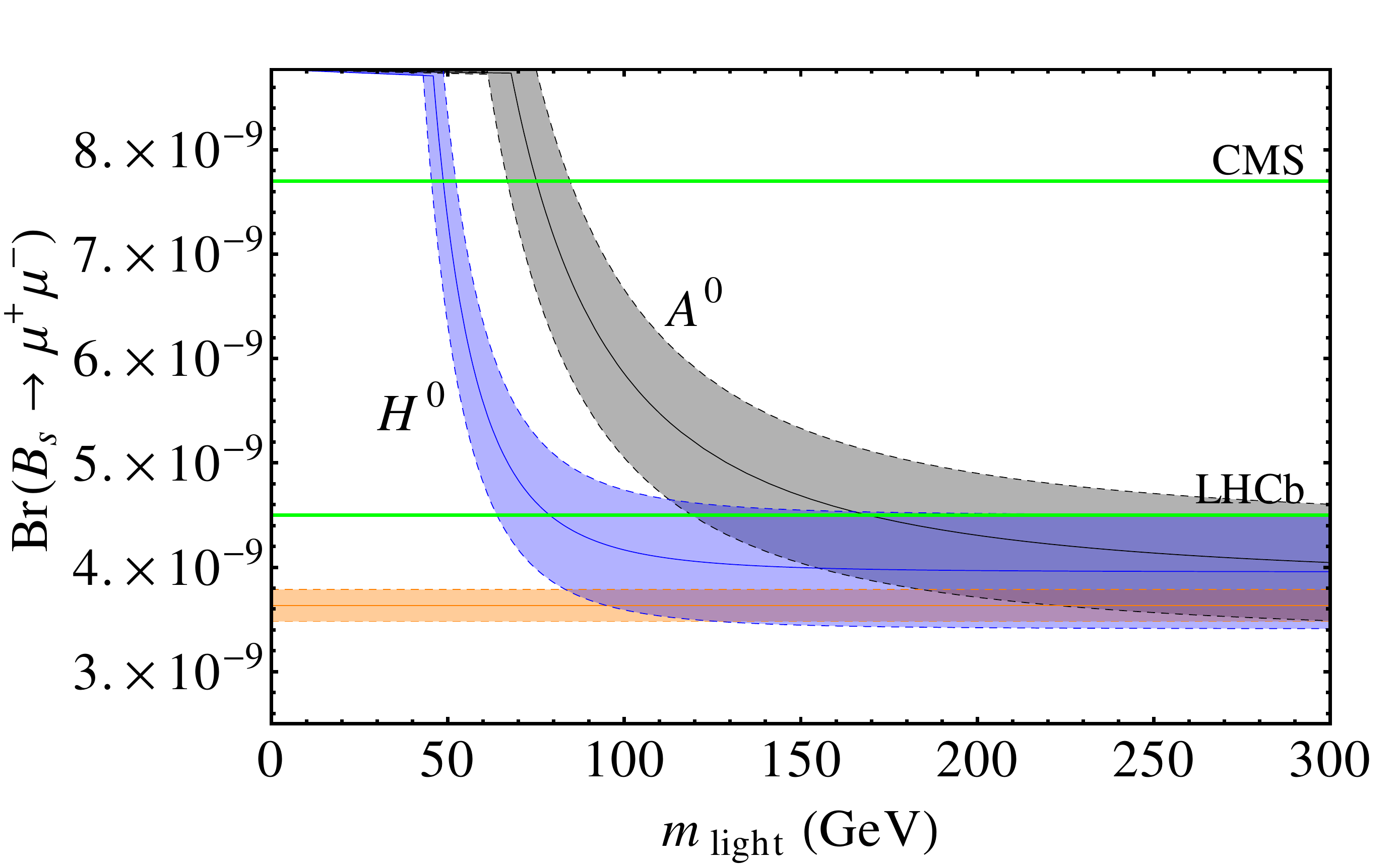}	
 \caption{\small{\textit{The $B_s\to\mu^+\mu^-$ branching ratio as a function of the $H^0$ and $A^0$ masses for the MFV coefficients $\delta_1=\lambda_0=1$. The thin horizontal (green) bands indicate the recent experimental upper bound values, the broad horizontal (orange) band indicates the 1$\sigma$ SM prediction, the lower (blue) curved band shows the 1$\sigma$ prediction for $m_{H^0}\ll m_{A^0}=400$ GeV, while the upper (grey) curved band corresponds to the analogue prediction for $m_{A^0}\ll m_{H^0}=400$ GeV. The theoretical value of the SM and the experimental bounds are given in Tab.\ref{numerics2}.}}}\label{fig:bto2mu}\end{figure}

From eq. \eqref{branching} we expect quite different behaviours depending on whether the lightest spin-0 particle is $A^0$ or $H^0$. Indeed, parity implies that the SM $Q_A$ operator only interferes with the $A^0$-induced $Q_P$. Consequently, the $B_s\to\mu^+\mu^-$ branching ratio is linear in $C_P$ but quadratic in $C_S$. That explains why, in figure \ref{fig:bto2mu}, the contribution of a 2HDM with $A^0$ the lightest flavour-violating spin-0 particle is more important

To summarize this section on flavour physics, let us enphasize once more that within MFV, the expansion coefficients in eqs. \eqref{ZdMFV} and \eqref{ZuMFV} are $\mathcal{O}(1)$ to fulfill the naturalness condition expressed hereabove. To display the maximal effect of a custodial 2HDM on K and B physics, we have simply taken $\delta_1=\lambda_0=1$ in Fig.\ref{neutralmesons} and \ref{fig:bto2mu}. With these natural values, the $B_s$ mixing provides a lower bound around 150 GeV for the lightest $H^0$ or $A^0$. However, given the theoretical uncertainties, if these coefficients are slightly smaller (say 1/2), flavour physics alone would still allow for $m_{H^0,A^0}$ around 100 GeV. Yet, in this case the constraint from $B_s\to\mu^+\mu^-$ becomes weaker than the one displayed in Fig.\ref{fig:bto2mu} and almost no deviation from the SM can be expected.

\begin{table}[htbp]
		\begin{center}\begin{small}\begin{tabular}{|c|c|cccc}
		\hline SM predictions	&	Measurements\\
		\hline
		$|\epsilon_K|_{SM}= 1.82(29)\times 10^{-3}$  	&$|\epsilon_K|_{exp} = 2.228(11)\times10^{-3}$ \cite{Nakamura:2010zzi}\\
		$(\Delta M_s)_{SM}=119.1(16.6) \times 10^{-13}$ GeV	&$(\Delta M_s)_{exp} =117.0(0.8)\times 10^{-13}$ GeV \cite{Nakamura:2010zzi}\\
		$\mathcal{B}(B_s\to\mu^+\mu^-)_{SM} = (3.6 \pm 0.2 )\times 10^{-9}$ &$\mathcal{B}(B_s\to\mu^+\mu^-)_{exp} \left\{ \begin{array}{cc}<7.7 \times 10^{-9} \mbox{ at }95\% \mbox{ CL } \cite{BsmumuCMS}\\< 4.5 \times 10^{-9} \mbox{ at }95\% \mbox{ CL } \cite{Aaij:2012ac}\end{array}\right.$ \\
		\hline
		\end{tabular}\end{small}\end{center}
\caption{\small{\textit{Theoretical and experimental values of flavour physics observables. Our theory predictions are consistent with \cite{Charles:2004jd}. For the $\epsilon_K$ parameter, we also rely on the new inputs given in \cite{Brod:2011ty}.}}}\label{numerics2}\end{table}

\section{Two-photon signal(s) at the LHC}

Recently, the ATLAS \cite{Collaboration:2012sk} and CMS \cite{Chatrchyan:2012tw} LHC experiments at CERN have reported an excess of events in the two-photon invariant mass spectrum at around 125 GeV. The possibility that the $A^0$ or $H^0$ present  in a 2HDM is responsible for such a signal has already been considered in \cite{Burdman:2011ki} and \cite{Ferreira:2012my}, respectively. Let us briefly (re)consider these possibilities in the context of our custodial 2HDM characterized by eqs. \eqref{phi1} and \eqref{phi2}.

In the SM, the $\Phi_1$ doublet alone is responsible for the gauge boson and matter particle masses ($M_W = \frac{gv}{2}$ and $m_t = y_t \frac{v}{\sqrt 2}$) \cite{Weinberg:1967tq}. As a consequence, $h^0$ is both vecto and fermiophilic and the dominant contributions to the diphoton events are due to top and W loops. In the 2HDM advocated here, the $H^0$ and $A^0$ are vectophobic (i.e. $g_{HVV} = g_{AVV} = 0$ with $V=W^{\pm},Z^0$) and only the top contributes. 

The number of events in the diphoton invariant mass spectrum is proportional to the production cross-section times the decay branching ratio. Remarkably, the ratio normalized to the SM rate
\begin{equation}
R=\frac{\sigma\times\mathcal{B}(H^0,A^0\to\gamma\gamma)}{\sigma\times\mathcal{B}(h^0\to\gamma\gamma)^{SM}}\label{ratio}
\end{equation} 

\noindent is rather sizeable and quite stable for spin-0 particles with a mass running from 0 to 125 GeV
\begin{equation}
R_{H^0/h^0}(m_{H^0}=0\to125 \mbox{ GeV}) = (0.12\to0.08) \frac{(\upsilon_0+\upsilon_1y^2_t)^4}{(\delta_0+\delta_1y^2_t)^2}\label{ratioA}
\end{equation}     

\begin{equation}
R_{A^0/h^0}(m_{A^0}=0\to125 \mbox{ GeV}) = (0.59\to0.44) \frac{(\upsilon_0+\upsilon_1y^2_t)^4}{(\delta_0+\delta_1y^2_t)^2},\label{ratioH}
\end{equation}  

\noindent if the production is dominated by gluon-gluon fusion via a top quark loop and in the limit where the total decay widths are dominated by the $b\bar b$ final state. (Note however that in the SM the $WW^*$ and $ZZ^*$ decays contribute to approximatively $25\%$ of the total width at 125 GeV.) This striking behaviour of the ratio R as a function of the (pseudo)scalar mass is due to the fact that the two-gluon and the two-photon couplings of a light spin-0 particle are determined by the so-called axial and scale anomalies. The corresponding effective Lagrangians describing these anomalies for $m_{H^0,A^0}\ll2M_W,2m_t$ are \cite{Kniehl:1995tn}
\begin{eqnarray}
\mathcal{L}_{(H^0,A^0)\gamma\gamma}&=&\frac{\alpha_{em}}{2\pi}\frac{1}{v}\{c_{\gamma\gamma}(+)H^0F^{\mu\nu}F_{\mu\nu}+c_{\gamma\gamma}(-)A^0F^{\mu\nu}\tilde F_{\mu\nu}\}\\
\mathcal{L}_{(H^0,A^0)gg}&=&\frac{\alpha_{s}}{12\pi}\frac{1}{v}\{c_{gg}(+)H^0G^{a\mu\nu}G^a_{\mu\nu}+c_{gg}(-)A^0G^{a\mu\nu}\tilde G^a_{\mu\nu}\}
\end{eqnarray}

\noindent with $\mathcal{O}(1)$ $c$-coefficients given by
\begin{eqnarray}
c_{\gamma\gamma}(+)=\frac{N_c}{3}q^2_t-\frac{7}{4}\hspace{0.5cm}&;&\hspace{0.5cm}c_{\gamma\gamma}(-)=\frac{N_c}{4}q^2_t\label{cgamma}\\
c_{gg}(+)=1\hspace{0.5cm}&;&\hspace{0.5cm}c_{gg}(-)=\frac{3}{2}.
\end{eqnarray}

\noindent In eq. \eqref{cgamma}, $q_t=\frac{2}{3}$ is the electric charge of the top quark while the second term of $c_{\gamma\gamma}(+)$ comes from the W-loop which is absent in the case of a vectophobic scalar $H^0$. 
\begin{figure}[htbp]
\centering\includegraphics[scale=0.31]{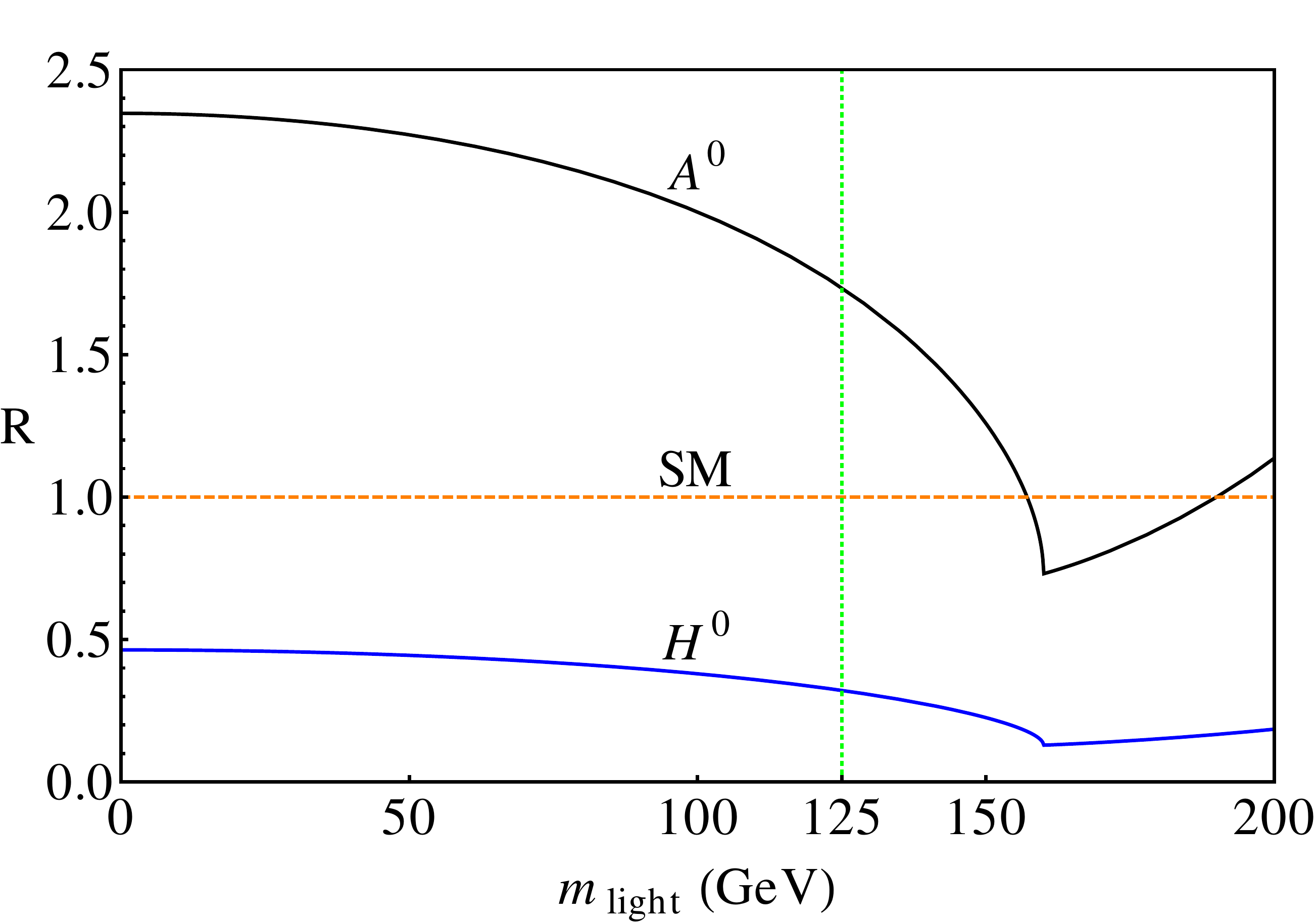}	
 \caption{\small{\textit{The ratio $R$ defined in eq. \eqref{ratio} as a function of  the $H^0$ and $A^0$ masses if the MFV coefficients are equal to one. The upper (black) curve corresponds to the case where $A^0$ is the lightest non-SM Higgs boson while the lower (blue) one corresponds to  the case where $H^0$ is the lightest non-SM Higgs boson.}}}\label{fig:hto2photons}\end{figure}

\noindent In figure \ref{fig:hto2photons}, the analytical expressions (explicitly given in \cite{Burdman:2011ki}) have been used to plot the ratio R as a function of the (pseudo)scalar mass. An interesting feature of our 2HDM is that the ratio R is of order one because of the naturalness principle of MFV ($\delta_i,\upsilon_i\approx 1$). In the scalar case (i.e. $H^0$-dominated), the number of events is expected to be smaller than in the SM even for large values of the MFV coefficients. The pseudoscalar case (i.e. $A^0$-dominated), on the contrary, is quite compatible with $R=1$ at 125 GeV for natural values of the $\upsilon_i$ and $\delta_i$ coefficients. It would also be able to account for a possible excess with respect to the SM expectation.

Two possible custodial scenarios with a light vectophobic $A^0(H^0)$ are thus within the reach of the present LHC data. The first one would correspond to the following mass hierarchy
\begin{equation}
m_{A^0(H^0)}<m_{ H^0(A^0)}\approx m_{H^{\pm}}<m_{h^0}\label{hierar1}
\end{equation}

\noindent with the $h^0$ mass in any case above the $A^0A^0(H^0H^0)$ threshold to avoid the LHC bounds on a heavy SM-like Higgs boson, and with a suitable mass-splitting between the custodian $A^0(H^0)$ and $H^{\pm}$ to satisfy the bounds from electroweak precision data. The second one, more consistent with respect to the custodial symmetry, would correspond to 
\begin{equation}
m_{h^0},m_{A^0(H^0)}<m_{ H^0(A^0)}= m_{H^{\pm}}\label{hierar2}
\end{equation}

\noindent with now two light resonances to be seen in the diphoton invariant mass spectrum. Needless to say that the forthcoming LHC results on Vector-boson fusion at the production level and on Vector-Vector final states at the decay level will be critical for these vectophobic scenarios. In particular, any excess from VV production or decay at around 125 GeV would rule out the scenario in eq. \eqref{hierar1} but not the one in eq. \eqref{hierar2} if the custodian $A^0(H^0)$ and $H^{\pm}$ are now sufficiently degenerated in mass to allow a light SM-like $h^0$ compatible with the $\rho$ parameter.

If future LHC data on $B_s\to\mu^+\mu^-$ and two-photon signals appear to be compatible with a single SM-like Higgs boson, the case where all new bosons are considerably heavier than the SM-like would still be allowed in the vectophobic 2HDM. Such a scenario corresponds to the usual decoupling regime,

\begin{equation}
m_{h^0}\ll m_{ H^0,A^0,H^{\pm}}.\label{decoupling}
\end{equation}

\section{Conclusion}

In this paper, we have considered a vectophobic 2HDM with a minimal violation of the flavour and custodial symmetries accidentally present in the SM.

On the one hand, the $B_s$ system provides us with the strongest indirect constraint on a light flavour-violating (pseudo)scalar. On the other hand, a direct way to test the model proposed in this work is the diphoton invariant mass spectrum at the LHC. In particular, a light custodian pseudoscalar $A^0$ could even allow for a two-photon excess with respect to the SM expectation. However, the $A^0$ and $H^0$ particles being vectophobic, any evidence of $W^+W^-$ or $Z^0Z^0$ gauge boson contributions at the production or decay level would also require a light SM-like scalar, namely a second diphoton signal.
Finally, let us underline that other channels could provide interesting signatures at the LHC. In particular, the allowed $H^0$-$A^0$-$Z^0$ coupling might induce a sizeable contribution to the $Z(\ell\bar{\ell})b\bar b$ cross-section as already enphasized in \cite{deVisscher:2009zb}.
\section*{Acknowledgments}
We would like to thank Fabio Maltoni and Christopher Smith for very useful discussions. The work of EC is supported by the National Fund for Scientific Research (F.R.S.-FNRS) under the FRIA grant. 

\bibliographystyle{elsarticle-num}
\bibliography{paperbibli}

\end{document}